\documentclass[useAMS,usenatbib]{mn2e}

\usepackage[psamsfonts]{amssymb}
\usepackage[dvips]{graphicx}
\usepackage{graphicx}
\usepackage{amsmath,alltt}                                                                                                                          
\usepackage{multirow}
\usepackage{rotating}
\usepackage{lscape}
\usepackage{tablefootnote}
\usepackage{hyperref}
\title[The Early Evolution of Star Clusters]{The Early Evolution of Star Clusters in Compressive and Extensive Tidal Fields}
\author[Webb, Patel, \& Vesperini]{Jeremy J. Webb, Saahil S. Patel \& Enrico Vesperini
\thanks{E-mail: jerjwebb@iu.edu (JW), evesperi@indiana.edu (EV)} \\
Department of Astronomy, Indiana University, Swain West, 727 E. 3rd Street, IN 47405 Bloomington, USA}

\begin{document}

\pagerange{\pageref{firstpage}--\pageref{lastpage}} \pubyear{2017}

\maketitle

\label{firstpage}

\begin{abstract}

We present $N$-body simulations of star clusters that initially evolve within a strong compressive tidal field and then transition into extensive tidal fields of varying strengths. While subject to compressive tides, clusters can undergo significant heating due to two-body interactions and mass loss due to stellar evolution. When the cluster transitions into an extensive tidal field it is super-virialized, which leads to a rapid expansion and significant mass loss before the cluster reaches virial equilibrium. After the transition, clusters are significantly less massive, more extended and therefore more tidally filling than clusters which have spent their entire lifetime in a similar extensive tidal field. 

\end{abstract}


\section{Introduction} \label{intro}

The majority of studies of globular cluster dynamics focus on the dynamical evolution of clusters within the present-day gravitational potential field of their host galaxy. While appropriate for studies of long-term cluster evolution, as clusters do indeed spend the majority of their lifetime in such a potential, this assumption might be less satisfactory when modelling a cluster's very early evolution. When the host galaxy is in the assembly stage, the surrounding tidal field experienced by newly formed star clusters is likely to be more complex than the present day tidal field \citep[see e.g.][]{kravtsov05, trenti15, kruijssen15, ricotti16, renaud17, li17}.

Only recently have studies begun taking into consideration that a cluster's formation environment can affect its early evolution and that the external tidal field undergoes significant evolution as the host galaxy forms and evolves \citep{renaud09, rieder13, miholics14, madrid14, renaud15, bianchini15, miholics16, renaud17}. \citet{renaud17} recently demonstrated that at high redshifts clusters form in regions where the net tidal field is compressive. A local compressive tidal field can be due to the superposition of the potentials of galaxies that are in the process of being assembled into a central host galaxy, the presence of a central bar or spiral arms \citep{martinez17}, or due to the cluster forming in the inner regions of a dwarf galaxy before being accreted \citep{bianchini15}. The role of these factors need to be considered before the properties of present-day globular clusters can be used to determine how clusters form and to constrain properties of their host galaxy.

Using $N$-body simulations to study the long-term dynamics of equal-mass star clusters in compressive tidal fields, \citet{bianchini15} found that evolving for many relaxation times in a compressive tidal field effectively heats a cluster as high velocities stars which would normally escape a cluster in isolation remain bound. While the subsequent transition from a tidally-compressive environment to isolation causes significant cluster expansion, the authors ultimately concluded that the expansion was not sufficiently strong enough to produce any significant structural or kinematic differences from a cluster that had spent its entire lifetime in isolation. Other studies (see e.g. \citet{hills80}, \citet{boily03,boily03b}, \citet{baumgardt07}) have explored the effects of rapidly removing an external potential on cluster evolution, where the external potential was due to leftover gas that was not converted into stars. However in the case of gas expulsion, the potential being removed is several orders of magnitude weaker than the compressive tidal fields considered by \citet{bianchini15}.

Building on the pioneering work done by \citet{bianchini15}, here we focus instead on the early evolution of clusters undergoing stellar evolution that have not experienced any significant tidal stripping and spend only a fraction of their initial half-mass relaxation time in a compressive tidal field. We specifically follow the dynamics of a cluster evolving in a compressive tidal field during the first 300 Myr of its life. During this phase the cluster loses mass due to the evolution of massive stars; the effects of mass loss due to stellar evolution have been explored in detail for clusters evolving in an extensive tidal field and have been shown to lead to an early cluster expansion and in some cases to a rapid dissolution (see e.g. \citealt{chernoff90, fukushige95}; see also \citealt{vesperini10} and references therein). Given the possibility that a cluster's early evolution occurs in a compressive tidal field it is important to explore how this field can affect the cluster's response to early evolutionary processes and how the properties set in the primordial tidal field determine the cluster's subsequent evolution after its transition to an extensive tidal field.

The outline of this Letter is the following: we present the full suite of initial conditions used for the $N$-body models used in this study in Section \ref{s_nbody}. In Section \ref{results} we present our results, focussing our attention on the evolution of the cluster's total mass and its structural properties. We discuss and summarize our findings in Section \ref{s_discussion}.

\section{N-body models} \label{s_nbody}

To model star clusters under the influence of compressive tides, we make use of the direct $N$-body code NBODY6 \citep{aarseth03}. Initially the distribution of stellar positions and velocities follow that of a Plummer sphere \citep{plummer11} out to a cut-off radius of 10 times the cluster's initial half-mass radius $r_{m,i}$ and star velocities are set so that the cluster is in virial equilibrium with the surrounding tidal field. To study how clusters with different initial masses and sizes evolve in compressive tidal fields, we consider two systems with a total initial number of stars equal to 64,000 and 150,000 and initial half-mass radii $r_{m,i}$ equalling 3 pc and 6 pc. For each model, we adopt a \citet{kroupa01} initial mass function (IMF) to generate the initial distribution of stellar masses between 0.1 and 100 $M_\odot$. The evolution of individual stars, all assumed to have a metallicity of 0.001, follows the stellar evolution algorithm of \citet{hurley00}. 

To setup a strong external compressive tidal field we follow \citet{bianchini15} who, motivated by observations of dwarf galaxies \citep{mcconnachie12},
places their model clusters at the centre of a Plummer potential with a mass of $10^8 M_\odot$ and a scale radius 100 pc. Since we are primarily interested in cases where clusters spend a relatively short period of time in a compressive tidal field, analogous to clusters which form in one of the early dwarf galaxies that contribute to assembling a central host galaxy or that form via a large-scale galaxy merger \citep{renaud09, renaud17}, we only consider tidal field transitions that occur at early times. More specifically, we allow each model cluster to evolve in a compressive tidal field for 300 Myr. Each of our models spend only a fraction of their initial half-mass relaxation time in the compressive tidal field and are therefore dynamically young at the time of transition.


To study the effect that transitioning to an extensive tidal field will have on a cluster, we then move each model cluster to a range of extensive tidal fields for an additional 700 Myr. Extensive tidal fields are setup by placing the cluster in a Milky Way-like tidal field at galactocentric distances of 10, 15, 20, 50 and 80 kpc. The Milky Way-like tidal field is made up of a $1.5 \times 10^{10} M_{\odot}$ point-mass bulge, a $5 \times 10^{10} M_{\odot}$ \citet{miyamoto75} disk (with $a=4.5\,$ kpc and $b=0.5\,$ kpc), and a logarithmic halo potential \citep{xue08} that is scaled so the circular velocity at $8.5$ kpc equals 220 km/s. Model clusters are assumed to orbit in the plane of the disk. For comparison purposes, the evolution of model clusters that spend their entire lifetimes in a Milky Way-like tidal field were also evolved for 1 Gyr.


\section{Results} \label{results}

\begin{figure}
\centering
\includegraphics[width=8.6cm]{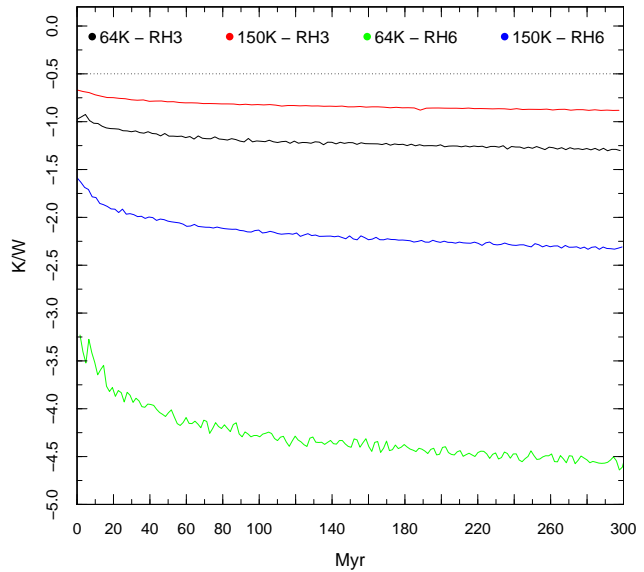}
\caption{Evolution of the ratio of kinetic energy to internal potential energy for model clusters with initially 64,000 stars and 150,000 stars and $r_{m,i} = $ 3 pc and 6 pc evolving in a strong compressive field (model clusters marked in the legend).
}
\label{fig:kw}
\end{figure}

In order to illustrate the effects of the external tidal field on cluster properties we start our analysis by focussing on the ratio of total kinetic energy ($K$) to the cluster's internal potential energy ($W$) and present its evolution in Figure \ref{fig:kw}. The potential energy, $W$, does not include the contribution of the external tidal field. We only consider the potential energy of the cluster when calculating $\frac{K}{W}$ as this provides a measure of how each cluster will respond when the strong compressive field is removed and the cluster starts evolving in an extensive field. Figure \ref{fig:kw} clearly illustrates how beginning in virial equilibrium with the initial tidal field actually corresponds to the cluster being in a strongly super-virial state when the contribution of the tidal field is not included (see e.g. \citet{bianchini15} for further discussion). A significant expansion is therefore expected once the initial strong compressive tidal field is removed. Neutron stars and black holes are given velocity kicks upon formation, with a velocity dispersion of 190 km/s \citep{hansen97}: their contribution to $\frac{K}{W}$ is not included as they temporarily cause spikes in $K$ when they first form before quickly escaping the cluster.

As high-mass stars lose mass via stellar evolution, the expansion and loss of stars that typically occur for clusters in extensive tidal fields are reduced by the presence of the strong compressive tidal field. Instead, mass loss due to stellar evolution heats the cluster further which causes $\frac{K}{W}$ to become even more negative. After the early burst of mass loss due to the stellar evolution of the most massive stars, two-body relaxation and tidal heating will cause $\frac{K}{W}$ to continue decreasing albeit at a much slower rate. Hence clusters will remain super-virialized until they are taken out of the compressive environment. As expected and clearly illustrated in Figure \ref{fig:kw}, the effects of the strong compressive external tidal field are weaker for more massive and more compact clusters (see \citet{bianchini15}).

\subsection{Mass} \label{mass}

\begin{figure}
\centering
\includegraphics[width=8.6cm]{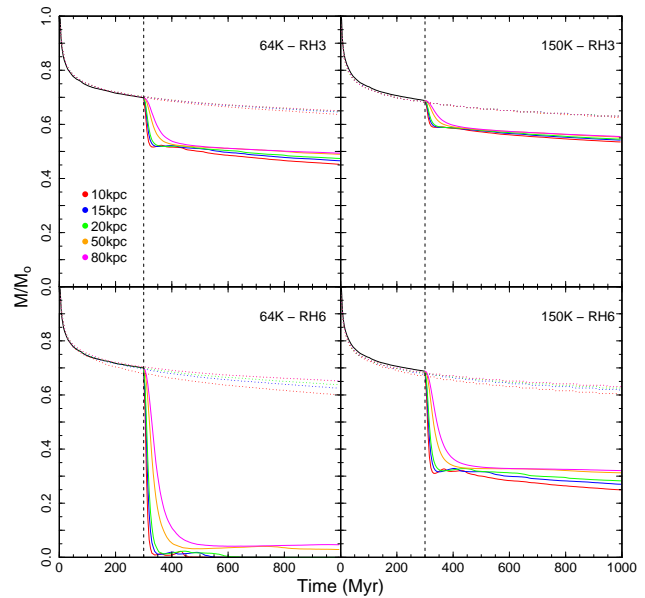}
\caption{Evolution of cluster mass for model clusters with initially 64,000 stars (left panels) and 150,000 stars (right panels) and $r_{m,i} = $ 3 pc (top panels) and 6 pc (bottom panels) initially evolving in a strong compressive field. After 300 Myr, models transition to Milky Way-like tidal field (galactocentric distances marked in the legend). Additional models that have always evolved in the Milky Way are shown as dotted lines.
}
\label{fig:mass}
\end{figure}

Figure \ref{fig:mass} shows the time evolution of cluster mass. During the compressive phase, only kicked dark remnants are energetic enough to escape the system so we consider all stars in the simulation when calculating global cluster properties. When clusters are in an extensive tidal field, we only consider stars within the cluster's tidal radius $r_t$. The initial drop in mass experienced by each model is due to mass loss associated with stellar evolution and the escape of dark remnants. Clusters that always evolve in extensive tidal fields undergo additional mass loss via the escape of stars that have been accelerated beyond the escape velocity of the cluster and tidal stripping. Clusters evolving in the strong compressive tidal field, on the other hand, do not do not suffer any significant additional mass loss during the initial 300 Myr.

As a consequence of their super-virial state, clusters that transition from a compressive to an extensive tidal field undergo significant expansion. There is also a burst of mass loss during the expansion as high-velocity stars no longer have to overcome a strong compressive tidal field to escape the cluster. The main factors which determine how much mass is lost during the transition are the degree of super-virial heating reached during the compressive phase (traced by $\frac{K}{W}$, which, as discussed in the previous section, depends on the size and mass of the cluster) and the strength of the extensive tidal field the cluster transitions to. High-mass clusters are heated less and have higher escape velocities than low-mass clusters, hence they will retain a higher fraction of stars during the transition. Similarily, compact clusters retain more mass during the transition because they are heated less than more extended clusters. 

The stronger the extensive field that the cluster transitions to (i.e the smaller the $R_{gc}$), the more mass is lost as the tidal field is now causing additional stars to escape via tidal stripping. Less compact clusters moving at closer galactocentric distances will suffer a rapid additional mass loss due to tidal stripping such that the initially low-mass models quickly reach dissolution. All compact and high-mass clusters, on the other hand, survive the transition and will continue to evolve normally until dissolution. 

\subsection{Structure} \label{structure}

\begin{figure}
\centering
\includegraphics[width=8.8cm]{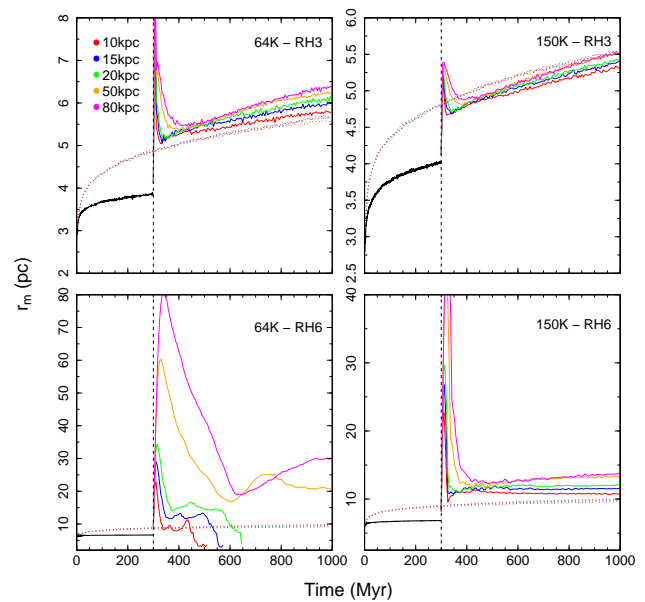}
\caption{Evolution of cluster half-mass radius for model clusters with initially 64,000 stars (left panels) and 150,000 stars (right panels) and $r_{m,i} = $ 3 pc (top panels) and 6 pc (bottom panels) initially evolving in a strong compressive field. After 300 Myr, models transition to Milky Way-like tidal field (galactocentric distances marked in the legend). Additional models that have always evolved in the Milky Way are shown as dotted lines.
}
\label{fig:rm}
\end{figure}

Figure \ref{fig:rm} shows the time evolution of the half-mass radius for all the models and clearly demonstrates the differences in the structural evolution of clusters undergoing their early evolutionary phases in a strong compressive field and those which instead always evolve in a standard extensive field . The presence of a strong compressive external field significantly inhibits the early expansion triggered by the mass loss due to stellar evolution and this effect is stronger for initially more extended clusters which are more affected by the external potential. After the transition, high-velocity stars migrate outwards causing the transitioning clusters to quickly expand. The extent of the expansion depends on the degree to which a system is heated during the compressive phase; as anticipated in the previous discussion of the evolution of the virial ratio, less compact clusters are characterized by a stronger heating and undergo a more significant expansion. In all cases since a large fraction of stars escape the cluster as it expands, the post-transition expansion is followed by an abrupt decrease in $r_m$ until the cluster reaches virial equilibrium in the new extensive potential.  

\begin{figure}
\centering
\includegraphics[width=8.8cm]{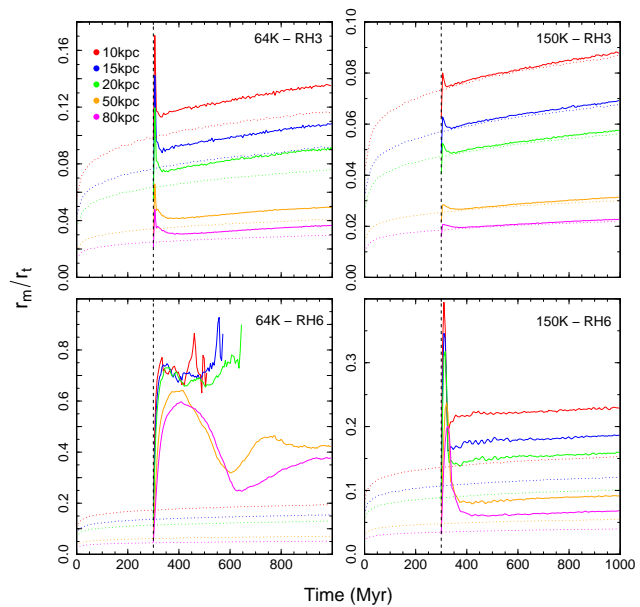}
\caption{Evolution of tidal filling factor $\frac{r_m}{r_t}$ for model clusters with initially 64,000 stars (left panels) and 150,000 stars (right panels) and $r_{m,i} = $ 3 pc (top panels) and 6 pc (bottom panels) initially evolving in a strong compressive field. After 300 Myr, models transition to Milky Way-like tidal field (galactocentric distances marked in the legend). Additional models that have always evolved in the Milky Way are shown as dotted lines.
}
\label{fig:rfill}
\end{figure}

In all cases, once each cluster transitions to its new $R_{gc}$ and reaches virial equilibrium it is larger than clusters which spend their entire lifetime at the same $R_{gc}$. However for the compact higher-mass clusters (model 150K-RH3) the difference is minimal (and $r_m$ is in this case even a bit smaller than that of the extensive counterpart). Models which transition to weak extensive tidal fields (larger $R_{gc}$) are significantly larger then their extensive counterparts as they have been able to expand without losing stars via tidal stripping. Since the transitioning clusters will also have lower masses, they will be characterized by lower densities and higher tidal filling factors ($\frac{r_m}{r_t}$). To illustrate this point, the evolution of each cluster's $\frac{r_m}{r_t}$ is shown in Figure \ref{fig:rfill} once it has reached the extensive tidal field.

Figure \ref{fig:rfill} demonstrates that clusters which initially evolve in a compressive tidal field will expand and lose enough mass after transitioning to an extensive tidal field such that they will be more tidally filling than clusters which always evolve in an extensive tidal field. However it should be noted that the tidal filling factors of the compact higher-mass clusters (model 150K-RH3) are only slightly larger than clusters which have always evolved in an extensive tidal field since they are heated less during the compressive phase and lose less mass during the transition to an extensive tidal field.

\section{Discussion and Summary}\label{s_discussion}

Our $N$-body simulations have shown that an early evolutionary phase in a compressive tidal field before transitioning into a extensive tidal field can significantly alter the mass and structure of a star cluster. Specifically, during the initial evolution in a strong compressive field a cluster will be characterized by a ratio of kinetic energy to internal potential energy corresponding to a supervirial state which leads to a strong expansion and mass loss once the effects of the strong compressive external potential are removed. 

Compared to clusters that always evolve in an extensive tidal field, clusters which transition from compressive to extensive tidal fields are less massive and larger (hence less dense) and thus, in general, characterized by larger tidal filling factors ($r_m/r_t$) at a given $R_{gc}$. In some of the extended lower-mass cluster cases we have explored, the expansion following the transition in the extensive tidal field resulted in complete cluster dissolution within 1 Gyr.

Taking into consideration the additional mass loss and expansion following the early evolution in a compressive tidal field, this evolutionary step can have several important implications for the evolution of the internal properties of individual clusters as well as for the global properties of globular cluster systems. Our simulations show that low-mass clusters may not survive the expansion following the transition from the compressive to the extensive tidal field (or in any case lose a very large fraction of their initial mass) even if the extensive fields we have considered are the relatively weak fields of the Milky Way at galactocentric distances larger than 10 kpc. At those distances clusters always evolving in an extensive tidal field do not, in general, suffer significant mass loss. The disruption of low-mass clusters and the larger amount of mass loss for surviving clusters associated with the process we have explored in this Letter may thus play a key role in shaping the globular cluster system mass function even at large galactocentric distances.  

We point out that the mass loss occurring during the transition event would not preferentially remove low-mass stars and would thus not change the slope of the low-mass main sequence stellar mass function observed today (unless the cluster was characterized by an extreme primordial mass segregation already involving stars less massive than approximately 0.8 $M_{\odot}$). Such an early episode of mass loss would therefore leave no fingerprint on the stellar mass function and the mass lost by a cluster that can be inferred from its present-day mass function would represent a lower limit \citep[see e.g.][]{vesperini97, webb15}. In the context of the study of multiple-population clusters, chemically enriched stars are predicted to form in the cluster inner regions \citep[see e.g.][]{dercole08} and would therefore not be affected by this early mass loss process. Hence transitioning from a compressive to an extensive tidal field may help drive the evolution of the fraction of chemically enriched stars towards the values currently observed in old globular clusters. 

The rapid expansion associated with clusters transitioning into an extensive tidal field must be considered when linking the initial distribution of cluster sizes to that of present day clusters. The expansion triggered by the tidal field transition discussed here offers a possible explanation for how initially compact primordial clusters can expand to reach their present-day sizes. The creation of clusters with $\frac{r_m}{r_t} > 0.1$ at $R_{gc} > 50$ kpc is of particular interest, as it suggests that the dynamical evolution immediately following the transition from compressive to extensive tidal fields could produce extended tidally filling clusters at large $R_{gc}$ \citep{baumgardt10, zonoozi11, zonoozi14}.

Ultimately, evolving in a compressive tidal field (even for a short period of time) represents an important phase in a cluster's dynamical history. The mass loss and expansion associated with transitioning into an extensive tidal field will have implications on the long term evolution of globular clusters and their present-day properties. Exploring a wider range of compressive field strengths and transition times will help shed further light on this early phase of dynamical evolution and help unravel how globular clusters form and what their initial conditions are at birth.

\section*{Acknowledgements}

We would like to thank the referee, Florent Renaud, for his many helpful suggestions to improve the manuscript. This work was made possible in part by Lilly Endowment, Inc., through its support for the Indiana University Pervasive Technology Institute, and in part by the Indiana METACyt Initiative. The Indiana METACyt Initiative at IU is also supported in part by Lilly Endowment, Inc.

\bsp

\label{lastpage}

\end{document}